# The Heun's functions as a modern powerful tool for research in different scientific domains

P. P. Fiziev

*Abstract:* In this rather popular paper we stress some basics of the theory and the huge amount of applications of the Heun's functions, as well as the still existing basic unsolved problems. The point is to attract more attention to this modern powerful tool for research in different scientific domains.

## Introduction

As a tool of the 21st century for solving theoretical, practical and mathematical problems in all scientific areas, the Heun's functions are a universal method for treatment of a vast variety of phenomena in complicated systems of different kind: in solid state physics, crystalline materials, Helium atom, water molecule, the Stark effect, graphene, in celestial mechanics, quantum mechanics, quantum optics, quantum field theory, atomic and nuclear physics, heavy ion physics, hydrodynamics, atmosphere physics, gravitational physics, black holes, compact stars, and especially in extremely urgent and expensive search for gravitational waves, astrophysics, cosmology, biophysics, studies of the genome structure, mathematical chemistry, economic and financial problems, etc. This wide area of application is a result of the general type of the Heun's differential equation that properly describes processes in all scientific areas.

## Basic historical remarks and developments

In the 21st century the Heun's equations represent what the hypergeometric equations represented in the 20th century. The first ones were introduced by the modest German mathematician Karl Heun (1859-1929) in his seminal paper [1], as a generalization of the hypergeometric equation studied by Gauss, Remann, Kummer and many others. The point was to construct the most general linear differential equation with four simple singular points, having in mind that the hypergeometric one owns only three of them.

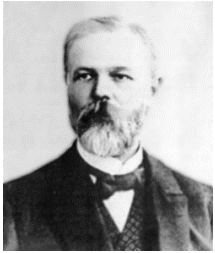

**The general Heun's differential equation:**

$$y'' + \left( \frac{\gamma}{z} + \frac{\delta}{z-1} + \frac{\varepsilon}{z-a} \right) y' + \frac{(\alpha \beta z - q) y}{z(z-1)(z-a)} = 0$$

The result depends on some number of free parameters (a,q,α,β,γ,ε,) which have different values in different applications. The general theory of such differential equations turns to be extremely complicated and at present is far from being complete, despite the fact that at first glance the general equation looks not much sophisticated. As shown by the Russian mathematician Vladimir Ivanovich Smirnov in his PhD Thesis (Petersburg, 1918), the fundamental reason is that one can throw a circle through any three points in the complex z-plane, but it's impossible to do this having four arbitrary points. Coalescing two or more simple singular points in the above general Heun's equation, one obtains four special Hein's equations: confluent, bi-confluent, double-confluent, tri-confluent ones [2,3] which have a lot of applications.

The development of the theory of the Heun's functions was out of the scope of mathematicians for a long time and the present explosion of interest was caused one century after the pioneer paper by Karl Heun only under pressure of the rapidly increasing number of their applications.

It is not hard to explain this phenomenon, quite unusual for mathematical methods and especially for mathematical and theoretical physics.

In any scientific area, studying the Nature, we first consider the simplest states of the corresponding specific systems (physical, chemical, biological, social, financial, etc.). Obviously, such simplest states are the equilibrium ones (stable or unstable), and the system can stay in such states during an arbitrarily long time period, until some external factor takes it away. Usually, we meet systems in stable equilibrium, or near equilibrium states, and their theory is simple.

If the deviation of the state from equilibrium is small, one can linearize the dynamical problem. Then, for any real system one can use a simple math for description of the motion in the vicinity of the equilibrium state. It describes the well-known harmonic oscillations around the equilibrium state.

For larger deviations one is forced to consider more complicated anharmonic oscillations which lead to different complicated phenomena, up to the chaotic behavior of the corresponding dynamical systems.

Besides, we often have to study systems, the behavior of which is described as spreading of waves. Typical examples are the classical particle dynamics and the quantum particle wave dynamics in the same potential U(x) of some external force (electric, gravitational, etc.). In the thirties of the last century, when the quantum mechanics was born, many theoreticians were trying to solve in the framework of the new wave mechanics the simplest exactly solvable problems, known from classical mechanics.

The increasing collection of many concrete examples showed a simple picture, which is not well known to large scientific audience: if the classical problem with a given potential U(x) can be solved in terms of exponential functions (including sin and cos functions), then the corresponding quantum problem with the same potential U(x) can be solved exactly in terms of hypergeometric functions. Examples can be found in any good textbook on quantum mechanics. It took much more time for the next step, namely, the recognition of the fact that if the classical problem can be solved in terms of elliptic functions, then the corresponding quantum problem can be solved exactly in terms of the Heun's functions. For example, mathematical pendulum, or anharmonic oscillators with potential U(x)= ½ k $x^2 + bx^3 + c\,x^4$.

In 1965 the above result was generalized by Sergey Yurievich Slavyanov to its strongest form known at present: if the classical problem can be reduced to a solution of any of the six classes of the Paul Painlevé's equations, then the corresponding quantum problem can be solved in terms of the Heun's functions [3]. In principle, this statement solves a huge amount of problems in different applications of all scientific areas.

## Selected modern problems and achievements

The most frequently used Heun's function is the confluent one with three singular points, two of which are simple and the third is obtained by confluence of two simple ones at infinity. The physical reason is simple: usually we are considering wave problems (like Hydrogen atom, which is solved by hypergeometric functions) in which there exist two waves: one going to infinity and a reverse one coming from infinity. Mathematically, this means that the infinite point must be an irregular singular point. The corresponding confluent Heun's equation was written in [4] in the following simple form

$$H'' + \left(\alpha + \frac{\beta+1}{z} + \frac{\gamma+1}{z-1}\right) H' + \left(\frac{\mu}{z} + \frac{\nu}{z-1}\right) H = 0$$

In this article, the basic theory of many special solutions of this equation important for applications was developed. The new achievements were applied by many researchers for solution of different problems. As a result, only in 2014 this article was used and cited more than in 23 independent studies in different scientific areas. Such impact is not common in the field of mathematical physics and applied mathematics. It is an evidence of that the role of the Heyn's functions was finally recognized by the scientific community.

As an illustration, we present here some results in the theory of gravitational waves in general relativity. After fifty years of development of the topic by different authors, in [5] the famous Regge–Wheeler equation for gravitational waves in the Schwarzschild metric of heavy compact objects was for the first time solved exactly using the confluent Heun's functions. The obtained spectrum for different boundary conditions at the body surface is shown in Fig.2. It gives a unique possibility to "see" the event horizon of black holes (which correspond to the black points), if they exist in the Nature. The body of the same mass and different nature will have another spectrum (see the color lines in Fig 2) depending on the reflection coefficient of the body surface:

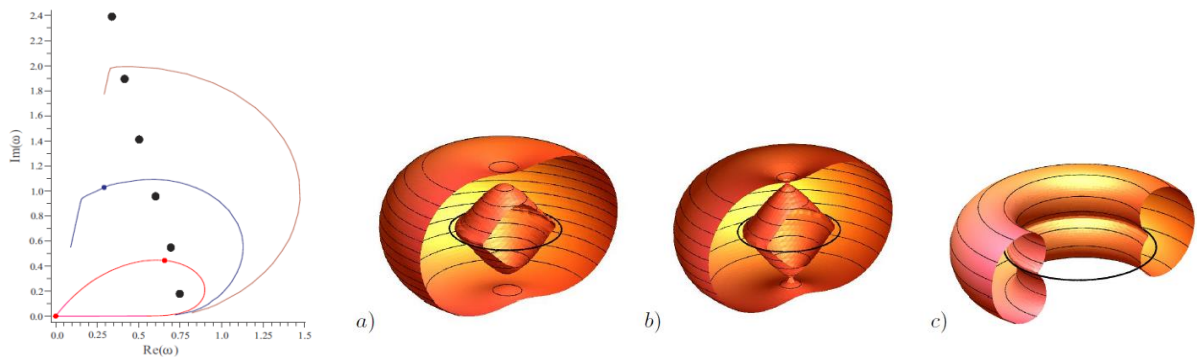

More complicated Teukolsky's master equations, which describe waves of different spin in rotating Kerr metric of heavy compact objects, were also solved exactly and used for the first time in [6]. The effect of bifurcation of ergosurface of the Kerr metric when the rotation parameter α is a) α < M – black hole , b) α = M – extreme black hole, and c) α > M – naked singularity (M being the mass of the Kerr solution) is shown in Fig.3.

The corresponding effects of the above bifurcation on the spectrum of electromagnetic waves in the Kerr metric are shown in Fig. 4 for real and for imaginary parts of several frequencies.

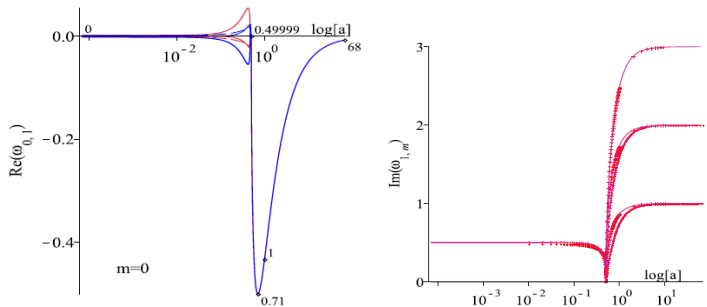

It is useful to mention also some part of the latest applications [7-14] of the Heun's functions.

**Unsolved problems and concluding remarks**

At the end, we wish to mark some topical problems for further applications of the Heun's functions.

The main unsolved problem is still the development of the general theory and the computational methods for the Heun's functions. Despite many efforts, at present we do not know the relations between different local solutions of the Heun's equation (the connection problem), the details of their monodromy group, the character of their irregular points, and so on. A special attention should be paid to the ambiguity of the functions of the Heun's functions and their branching points. Typically, they are ignored completely, leading to serious computational problems [15]. This restricts not only our general understanding of the Heun's functions, but is also a serious obstacle for development of the effective computational methods for their applications.

Still, the only computer package which is able to make analytical and numerical computations with the Heun's functions is Maple. The author is grateful to Dr. Edgardo Cheb-Terrab from MapleSoft for many years of fruitful collaboration and help in the usage of the Maple package for the Heun's functions, as well as for its successive development. One ought to stress that in the Maple 2015 there is an essential progress in the high precision calculations with the confluent Heun's functions. Nevertheless, there still remain many unsolved essential problems in Maple package for the Heun's functions.

One can expect a further intensive development of the theory, computational methods and applications of the Heun's functions in the nearest future. The richest collection of references on this subject can be found at the WEB address [7]. There is also information about our "The Heun Project" which is in development and search of financing for contacts, workshops, conferences on the topic, as well as for education and dissemination of the results. At present, a significant group of leading researchers of this subject all over the world participates in the project and makes efforts for its further development.